\documentclass{aa}  

\usepackage{keyval}
\usepackage{graphicx}
\usepackage{hyperref}

\begin{document}

	\title{Statistical study for binary star evolution in dense embedded clusters}
	
	%   \subtitle{}
	\titlerunning{BP in embedded cluster}
	\authorrunning{W. Wu et al.}
	\author{
		Wenjie Wu\inst{1},
		Pavel Kroupa\inst{1,2},
		Vikrant V. Jadhav\inst{2}}
	
	\institute{
		Helmholtz-Institut für Strahlen- und Kernphysik, Universität Bonn, Nussallee 14-16, D-53115 Bonn, Germany\\
		\email{wwu@astro.uni-bonn.de}
		\and
		Astronomical Institute, Faculty of Mathematics and Physics, Charles University, V Holešovičkách 2, CZ-180 00 Praha 8, Czech Republic}
	
	\date{Received MM DD, YYYY; accepted 23 January, 2026}

	\abstract
	{The dynamical evolution of binary populations in embedded star clusters shapes the statistical properties of binaries observed in the Galactic field. Accurately modelling this process requires resolving both early cluster dynamics and binary interactions.}
	{We aim to characterize the early dynamical evolution of primordial binaries in embedded clusters and identify the key parameters that govern binary survival and disruption.}
	{We perform a set of direct $N$-body simulations starting from 100\% primordial binaries in a time-varying gas potential of a gas-embedded cluster. To describe the evolution of binary orbital properties, we define empirical dynamical operators for period, binding energy, and mass ratio, and calibrate them across the simulated ensemble.}
	{The binding energy and orbital period evolve in a consistent, sigmoidal fashion. Their dynamical operators reveal that hard binaries heat the cluster and suppress wide binary formation, while a small residual population of soft binaries survives. The evolution of the mass-ratio distribution is less directly linked to dynamical processing and more shaped by internal processes such as stellar physics process in the pre-main-sequence phase. High-$q$ systems tend to be enhanced, while low-$q$ systems are prone to disruption.}
	{The binary evolution in clusters is primarily governed by binding energy and orbital period. Our model improves over previous parameterizations of the dynamical operator by allowing for the existence of wide binaries and incorporating the embedded cluster phase. For individual clusters, direct $N$-body modelling remains the only reliable approach. On Galactic scales, population synthesis methods based on the stellar dynamical operator approach developed here remain essential.}
	
	\keywords{Galaxies: star clusters: general -- (Galaxy:) open clusters and associations: general -- (Stars:) binaries: general -- Stars: kinematics and dynamics -- Methods: numerical -- Methods: statistical }
	
	\maketitle
	
\section{Introduction}

A key factor in dynamical modelling of stellar population is the characterization of the binary population (BP), which comprises the distributions of binary parameters such as primary mass, orbital period, eccentricity, and mass ratio. Numerous studies have shown that the dynamical evolution of star clusters is significantly influenced by the presence and properties of binary systems \citep[e.g.,][]{1975MNRAS.173..729H, 1995MNRAS.277.1491K, 1995MNRAS.277.1522K,2009MNRAS.397.1577P, 2011A&A...529A..92K}. In the context of stellar population synthesis, particularly when accounting for unresolved binaries, a reliable model for the BP is crucial for accurately reconstructing the stellar mass function \citep[e.g.,][]{1991MNRAS.251..293K,2011MNRAS.417.1702M,2024ApJ...966..169P,2025CoSka..55c.217K}.

It is widely accepted that the Galactic field stellar population originates predominantly from the dissolution of embedded star clusters that form in molecular cloud clumps  \citep[e.g.,][]{1995MNRAS.277.1491K,1995MNRAS.277.1522K,2003ARA&A..41...57L, 2010MNRAS.409L..54B, 2010RSPTA.368..713L, 2011IAUS..270..141K, 2011MNRAS.417.1702M, 2025CoSka..55c.217K}. Following the expulsion of residual gas, an embedded cluster rapidly loses its dynamical equilibrium, leading to the loss of a significant fraction of single and multiple stellar systems into the Galactic field. As shown in studies such as \citet{2001MNRAS.321..699K} and \citet{2002ASPC..285..141B}, the cluster undergoes a phase of rapid expansion followed by re-virialization, ultimately reaching a gas-free state. These post-gas-expulsion phases are less dynamically active than the deeply embedded phase, suggesting that the BP's dynamical evolution prior to gas expulsion warrants detailed investigation.

Direct observation of the BP, however, is challenging. Prior to gas expulsion, the residual gas and dust obscures stellar light, making direct observation nearly impossible. During the rapid expansion phase, stars become observable, and several techniques may be employed. For close binaries with semi-major axes $a < 10$ AU, spectroscopic monitoring and eclipsing binary analysis can be used to infer binary properties \citep[e.g.,][]{2012Sci...337..444S, 2019AJ....157..196K}, though these methods require long observational times and are less sensitive to low mass ratios. Interferometry, sparse aperture masking, high-resolution imaging, and astrometry can detect binaries at intermediate separations ($a \approx 1$–300 AU), depending on the distance to the system \citep[e.g.,][]{2014ApJS..215...15S, 2019ApJ...886...95D}. However, these techniques are often infeasible for distant star-forming regions. For systems with high mass ratios or wide separations, analyses using colour–magnitude diagrams (CMDs) and proper motion data are effective \citep[e.g.,][]{2020AJ....159...15T, 2021AJ....162..264J}, although such methods also have clear limitations. As a result, detailed modelling of the BP using $N$-body simulations becomes essential.

\citet{1995MNRAS.277.1491K,1995MNRAS.277.1522K} introduced the concept of inverse dynamical population synthesis by performing a series of direct $N$-body simulations with 200 primordial binaries using the \texttt{nbody5} code \citep{1999PASP..111.1333A}. In these calculations, the binaries were initially embedded in a bound stellar cluster, and their dynamical evolution was followed self-consistently.This work led to the formulation of the initial binary star distribution functions. The resulting changes in the BP invoked by the stellar-dynamical processing of the BP were subsequently encapsulated through operators acting on the corresponding distribution functions, as derived from the simulation outcomes \citep[see also][]{kroupa2002hab}.
Building upon this approach, \citet{2011MNRAS.417.1684M} extended the range of initial cluster parameters—such as the initial half-mass radius and cluster mass (up to $10^{3.5}~\mathrm{M_\odot}$, corresponding to approximately 6000 stars or 3000 binaries)—and investigated how these parameters influence the dynamical operators. These studies laid the groundwork for the development of the binary population synthesis code \texttt{BiPoS1} \citep{2012MNRAS.422.2246M, 2022MNRAS.510..413D}, which enables inference of binary populations on galactic scales.

Modelling the dynamical evolution of binary-rich, high-mass star-forming regions such as the Orion Nebula Cluster (ONC), which contains on the order of $10^4$ stars and brown dwarfs, remains a major challenge for direct $N$-body simulations. Moreover, the deeply embedded phase of open clusters likely had significantly higher masses than observed today. For instance, \citet{2001MNRAS.321..699K} \citep[see also][]{2025MNRAS.541.1753S} suggest that the initial mass in stars and brown dwarfs of the Pleiades was between 3000 and 4000~$\mathrm{M_\odot}$, compared to its present-day mass of roughly 800~$\mathrm{M_\odot}$. Since the dynamical evolution of binary systems is primarily driven by close encounters rather than by the overall cluster potential, a direct 
$N$-body method needs to be employed to quantify the dynamical processing of the BP \citep[see, e.g.,][]{2023LRCA....9....3S}. 

As discussed in \citet{2020MNRAS.497..536W}, direct $N$-body codes such as \texttt{nbody6} and its parallelized extension \texttt{nbody6++GPU} \citep{2015MNRAS.450.4070W} face significant performance bottlenecks when simulating binary-rich dense clusters, due to inefficient parallelization in handling close encounters. Consequently, previous applications of inverse dynamical population synthesis have been limited to clusters with masses below $10^{3.5}~\mathrm{M_\odot}$. \citet{2020MNRAS.497..536W} introduced a new direct $N$-body code, \texttt{PeTar}, which incorporates highly efficient parallel algorithms for integrating close systems. As a result, \texttt{PeTar} enables realistic and scalable modelling of binary-rich clusters in regimes previously inaccessible to other direct $N$-body frameworks.

The primary objective of this paper is to perform direct $N$-body simulations using \texttt{PeTar}, incorporating the most up-to-date findings from studies of the initial binary population (IBP). We consider an analytical model that isolates the dominant dynamical effects of residual gas during the deeply embedded phase of massive open clusters. The analysis is conducted within the framework of inverse dynamical population synthesis. We begin by examining the distributions of commonly studied binary parameters—orbital period, semi-major axis, eccentricity, and mass ratio. In addition, we extend the analysis to investigate the dependence of binary properties on primary mass and the binary fraction across different mass regimes.

The paper is organized as follows. Section~\ref{sec:method} outlines the initial conditions and $N$-body setup, and Section~\ref{sec:framwork} introduces the binary distribution functions and dynamical operators. The main results are presented in Section~\ref{sec:results}, followed by conclusions in Section~\ref{sec:conclusion}. Appendices provide supporting analyses of the period distribution, parameter fits and correlations, and a comparison with the analytic framework of \citet{2011MNRAS.417.1684M}.

\section{Method}
\label{sec:method}
In this study, we employ the state-of-the-art N-body code \texttt{PeTar} \citep{2020MNRAS.497..536W} to conduct N-body simulations. In this section, we describe the details of the setup of initial state  and the N-body integration.

\subsection{Setup of the initial configuration}
\label{sec:initial_setup}
The initial configurations for our model are generated using the \texttt{McLuster} code \citep{2011MNRAS.417.2300K}, which supports a wide range of initial conditions suitable for generating stellar distributions for $N$-body simulations. Below, we describe the specific configuration used in our study.

The cluster mass is set to $M_{\rm ecl} = 5000~\mathrm{M_\odot}$ (the actual value varies slightly; see details below), and the half-mass radius is set to $R_{\rm h,m} = 0.3~\mathrm{pc}$. This value is motivated by the empirical $M_{\rm ecl}$–$R_{\rm h}$ relation from \citet{2012MNRAS.422.2246M}, which yields $R_{\rm h} \approx 0.303~\mathrm{pc}$. In practice, this 1\% deviation cannot be implemented exactly due to the finite number of stars. To obtain statistically reliable results, we generate 100 realizations of the cluster using different random seeds, treating them as samples from the same model. Note, the setup procedure introduces deviations between the input and actual values of cluster parameters (e.g., $R_{\rm h, m}$, $M_{\rm ecl}$). We always use the actual ones calculated directly from the generated snapshot parameters.

\subsubsection{Initial mass function}
\label{sec:IMF}
We adopt the canonical initial mass function (IMF) $\xi(m)= {\rm d}N/{\rm d}m$ from \citet{2001MNRAS.322..231K}, where ${\rm d}N$ is the infinitesimal number of stars in the stellar mass range $m$ to $m+{\rm d}m$. It follows a two-part power-law form (up to a normalization constant):

\begin{equation}
\label{eq:IMF}
    \xi(m) \propto 
    \begin{cases}
    m^{-\alpha_1} & m_{\rm min} \le m < m_1, \\
    k_{\xi} m^{-\alpha_2} & m_1 \le m \le m_{\rm max},
    \end{cases}
\end{equation}
where $\alpha_1 = 1.3$, $\alpha_2 = 2.3$, $m_{\rm min} = 0.08~\mathrm{M_\odot}$, and $m_1 = 0.5~\mathrm{M_\odot}$. The constant $k_{\xi} = m_1^{\alpha_2 - \alpha_1}$ ensures continuity at the break mass $m_1$.

For the upper mass limit $m_{\rm max}$, we adopt the empirical $m_{\rm max}$–$M_{\rm ecl}$ relation from \citet{2006MNRAS.365.1333W} and \citet{2007MNRAS.375..855P} \citep[see also][]{2023A&A...670A.151Y} giving $m_{\rm max}=89.73~{\rm M_\odot}$. To suppress stochastic variations in the number of massive stars—which can significantly affect the dynamical evolution \citep[see e.g.][]{2021A&A...655A..71W}—we use the optimal sampling method introduced by \citet{2013pss5.book..115K}. In conventional random sampling, a uniform random variable $X \in (0, 1]$ is drawn, and the corresponding stellar mass $m_\star$ is found by solving
\[
\int_{m_{\rm min}}^{m_\star} \xi(m)\, \mathrm{d}m = X,
\]
i.e., $X$ represents the quantile corresponding to $m_\star$. In contrast, 
optimal sampling uses equal spaced values of $X \in (0, 1]$, with the number of grid points determined by $N = M_{\rm cl}/\langle m\rangle$, where $\langle m\rangle$ is the average stellar mass computed from Equation~\eqref{eq:IMF}. Thus optimal sampling ensures that the resulting mass array is deterministic and identical for a fixed cluster mass. A complete physical motivation and theoretical discussion can be found in \citet{2013pss5.book..115K,2026enap....2..173K,2026RAA....26b5003G}; here, we focus on its practical application rather than its underlying theory.

\subsubsection{Initial binary population}

We adopt a primordial binary fraction of 100\%. As discussed in \citet{1995MNRAS.277.1491K} and emphasized by \citet{2015ApJ...800...72T} and \citet{2025CoSka..55c.217K}, this assumption provides a consistent basis for exploring whether a single birth configuration can evolve into a binary population similar to that observed in the Galactic field and in star-forming regions. We follow the same hypothesis to ensure consistency across our model. The IBP is assigned separately for low-mass binaries (primary mass $< 5~\mathrm{M_\odot}$) and high-mass binaries (primary mass $> 5~\mathrm{M_\odot}$). Prior to pairing stars, a single mass-array is generated from the IMF (see section \ref{sec:IMF}). It is of crucial importance to draw only one primary and secondary component mass from the mass array in order to not affect the IMF.

For low-mass binaries, we adopt the prescription of \citet{2017MNRAS.471.2812B}. The primary mass $m_{\rm p}$ and secondary mass $m_{\rm s}$ are randomly and independently selected from the stellar mass-array, considering only stars with $m < 5~\mathrm{M_\odot}$ for pairing. For each cluster model the pair thus differ despite the models having the same mass array. The orbital period $P$ is drawn from the distribution in \citet{1995MNRAS.277.1522K} (equation~8 there), and the eccentricity $e$ is drawn from a thermal distribution ($\propto 2e$). The orbital phase and inclination are sampled from uniform distributions on $[0, 2\pi)$ and the unit sphere, respectively.

For high-mass binaries, motivated by the observed dynamical ejections of OB stars \citep{2017A&A...604A..22B,2018A&A...612A..74K,2019A&A...627A..57J}, a separate pairing procedure is adopted following \citet{2018A&A...612A..74K}, as implemented by Long Wang in the updated version of \texttt{McLuster} \citep[see][]{2019MNRAS.484.1843W}. The mass ratios are generated by choosing the secondary from the mass-array so that the mass-ratio distribution is consistent with the observational data\citep[see ][]{2012Sci...337..444S}, based on O-star observations in six nearby Galactic open clusters. The period and eccentricity are drawn from the empirical distributions provided in \citet{2015ApJ...805...92O} and \citet{2017MNRAS.471.2812B}.

We also adopt the pre-main-sequence eigenevolution model introduced by \citet{1995MNRAS.277.1522K} and updated in \citet{2017MNRAS.471.2812B}. This process modifies the mass ratio and eccentricity (and, accordingly, the period), accounting for mass transfer and circularization during the pre-main-sequence phase (spanning about first $10^5$ yr). 

The pre-main-sequence eigenevolution modifies the binary population on a system-by-system basis, transforming the birth binary population into the IBP. However, this procedure breaks the analytical properties of the original mass distribution and introduces stochasticity. As a result, even under optimal sampling, the final mass array differs slightly between realizations, and the actual cluster mass $M_{\rm ecl}$ may exceed the nominal input value.

\subsubsection{Initial phase-space distribution}

We initialize the positions and velocities of all binary centre-of-mass systems (c.m.s.) using a Plummer model \citep{1911MNRAS..71..460P}, treating each binary as a single particle. The positions and velocities of individual stars are then computed from their binary orbital parameters and the c.m. positions and velocities. In this study, we assume no primordial mass segregation.

After generating each cluster, we rescale the velocities (only the centre-of-mass velocities for binaries) to ensure virial equilibrium \citep[following ][]{1999CeMDA..73..127A} with the gas potential (see Section~\ref{sec:nbodycode}), such that the virial ratio is $Q = 0.5$. 

We note that our setup assumes the embedded cluster to be initially in a state of virial equilibrium. Recent studies on BPs have explored non-equilibrium initial conditions, such as fractal structures \citep[e.g.][]{2017MNRAS.465.2198D}, showing that the environment can significantly influence the dynamical evolution of BPs. However, as pointed out by \citet{2001MNRAS.321..699K}, the initial conditions of embedded clusters are subject to physical constraints. In reality, star formation proceeds over about a Myr until gas expulsion. Once a star forms, it either falls toward the cluster's potential minimum or is already on a "virial-equilibrium orbit" \citep[see also][]{2011MNRAS.410.2799M}. In either case, the accumulating stellar system is continuously virializing as new stars are added, since the crossing timescale of an embedded cluster (see section \ref{sec:timescales}) is much shorter than its formation timescale. Therefore, a young embedded cluster is most likely to be close to virial equilibrium.
\subsection{The $N$-body integration}
\label{sec:nbodycode}
 \texttt{PeTar} combines several advanced computational techniques to handle interactions across different distance scales: the Barnes-Hut tree algorithm \citep{1986Natur.324..446B} for long-range interactions, a fourth-order Hermite integrator with block time-step \citep[e.g., ][]{2003gnbs.book.....A} for intermediate-range interactions, and the slow-down algorithmic regularization method \citep{2020MNRAS.493.3398W} for short-range interactions.

\texttt{PeTar} integrates the \texttt{SSE/BSE} code \citep{2000MNRAS.315..543H, 2002MNRAS.329..897H, 2020A&A...639A..41B} to model both single and binary stellar evolution. The remnant mass is determined based on the rapid supernova scenario from \citet{2012ApJ...749...91F}. The calculations also include contributions from pair-instability \citep{2016A&A...594A..97B} and electron-capture supernovae \citep{2008ApJS..174..223B}. For all models, we assume a solar metallicity of \( Z = 0.02 \).

The \texttt{galpy} library \citep{2015ApJS..216...29B} has been integrated into \texttt{PeTar} as an extension to provide various types of external fields. To simplify the model, the model clusters orbit on a round orbit around a point-mass ($M_{\rm gal}=5\times 10^{10}~\mathrm{M_\odot}$) at a distance $D = 8.5$ kpc.

Gas is modelled as a static Plummer potential in the cluster-centric frame \citep[see e.g.,][]{2001MNRAS.321..699K,2013ApJ...764...29B}. The gas mass is set to the mass of the embedded cluster divided by the star formation efficiency (SFE). Following the previous $N$-body modelling and to reduce the number of parameters, we assume here $\text{SFE} = 1/3$. Both the stellar and gas potentials share the same Plummer radius. As discussed in \cite{2001MNRAS.321..699K}, the gas expulsion starts with a delay after the star formation phase, which is typically 0.6 Myr. Therefore we chose the end time of all simulations to be 0.6 Myr, because the thereafter emerging cluster expansion largely freezes the binary population. Gas expulsion drives substantial cluster expansion, the dynamical environment becomes less favourable for further binary processing beyond this point. Thus, limiting the integration time to 0.6~Myr allows us to isolate the effects of early cluster dynamics prior to the onset of gas-driven expansion.

We acknowledge that this model omits detailed star–gas interactions, such as dynamical friction with the gas and feedback-driven gas removal. In particular, gas dynamical friction may influence the evolution of high-mass binaries, potentially leading to tighter configurations or enhanced disruption rates. Recent studies have shown that gas interactions can harden binaries and extend the traditional soft–hard boundary \citep{2024ApJ...968...80R}, while gas-assisted capture provides an efficient channel for the formation of new binaries, especially among massive stars in gas-rich environments \citep{2023MNRAS.521..866R}.  Such effects are especially pronounced for massive or compact binaries.
 However, high-mass binaries constitute a minor fraction of the population in this study. As shown in Section~\ref{sec:mass_dependence_fbin}, most dynamical modifications occur among low-mass binaries. Therefore, we conclude that our gas model is sufficient to capture the dominant dynamical effects relevant to the scope of this work.

All computations were performed using 8 CPU cores with SIMD acceleration on the \texttt{MARVIN} cluster, hosted by the University of Bonn. By leveraging the HPC system's capability to run multiple samples in parallel, we completed the entire set of simulations within approximately one week.

\section{Binary distribution functions and the dynamical operator}
\label{sec:framwork}
We adopt the framework established by \citet{1995MNRAS.277.1491K} and \citet{2011MNRAS.417.1684M} to describe the dynamical evolution of binary star populations. In this section, we summarize the mathematical definitions and notations.

At any given time \( t \), the distribution of binary parameters—namely the absolute value of the binding energy \( E_{\rm b} \), orbital period \( P \), semi-major axis \( a \), eccentricity \( e \), primary mass \( m_{\rm p} \), secondary mass \( m_{\rm s} \), and mass ratio \( q = m_{\rm s}/m_{\rm p} \)—is described by the individual binary distribution function (BDF), denoted as \( \Phi_{\pi_i}^t \). Here $\pi_i$ represents one of  the binary parameters. The BDF characterizes the distribution of binary parameters within a population but is not equivalent to a probability density function (PDF); specifically, it is not normalized to unity. Instead, the integral of the BDF over the entire domain of \( \pi_i \) yields the total binary fraction \( f_{\rm bin}^t \) of the cluster at time \( t \):

\begin{equation}
	\int_{\pi_{i, \rm min}}^{\pi_{i, \rm max}} \Phi_{\pi_i}^{t}(\pi_i')\, \mathrm{d}\pi_i' = f_{\rm bin}^{t}.
\end{equation}

The time evolution of the BDF from an initial time \( t_0 \) to a later time \( t_{\rm end} \) is modelled using a dynamical operator \( \Omega_{\pi_i}(\pi_i ; \omega_j) \), where \( \omega_j \) is a set of fitting parameters that characterize the dynamical processes—such as \( \mathcal{E}_{\rm cut} \) and \( \mathcal{A} \), as defined in equation (20) of \citet{2011MNRAS.417.1684M}. The evolved BDF at time \( t_{\rm end} \) is given by:

\begin{equation}
	\Phi_{\pi_i}^{t_{\rm end}}(\pi_i; \omega_j) = \Omega_{\pi_i}(\pi_i ; \omega_j) \cdot \Phi_{\pi_i}^{t_0}(\pi_i).
\end{equation}

While a BDF shares some formal similarities with a PDF, the key distinction lies in normalization: a PDF integrates to unity, whereas a BDF integrates to the binary fraction. To empirically determine the BDF, we first estimate the empirical PDF and then scale it by the observed binary fraction. Consequently, the empirical dynamical operator can be directly computed as the ratio of the empirical BDFs at \( t_{\rm end} \) and \( t_0 \).

The functional form of the dynamical operator \( \Omega_{\pi_i} \) is chosen empirically, typically guided by the shape of the measured empirical operator. Physically, \( \Omega_{\pi_i}(\pi_i) \) can be interpreted as an importance function, representing the survival probability of binaries with parameter \( \pi_i \) after dynamical processing.

A closed-form expression for the initial BDF is generally not available, primarily due to the inclusion of pre-main-sequence eigenevolution. This process alters parameters such as eccentricity based on the birth pericentre, which itself depends on the birth period and eccentricity. Consequently, the final BDF, \( \Phi_{\pi_i}^{t_{\rm end}} \), is difficult to derive analytically, even when the form of \( \Omega_{\pi_i} \) is known. This complexity renders standard maximum likelihood estimation impractical for fitting \( \omega_j \). To overcome this challenge, we employ the Kullback–Leibler importance estimation procedure to determine the parameters \( \omega_j \). The theoretical basis for this method is detailed in, for example, \citet{NIPS2007_be83ab3e}. In brief, we maximize the following objective function:

\begin{equation}
	\label{eq:def_KLIEP_loss}
	\ell(\omega_i) = \frac{1}{N_{t_\mathrm{end}}} \sum_{k=1}^{N_{t_\mathrm{end}}} \log \left( \Omega_{\pi_i}(\pi_{i, t_\mathrm{end}}^{(k)} ; \omega_j) \right),
\end{equation}
subject to the normalization constraint:
\begin{equation}
	\label{eq:def_KLIEP_cond}
	\frac{1}{N_{t_0}} \sum_{k=1}^{N_{t_0}} \Omega_{\pi_i}(\pi_{i, t_0}^{(k)} ; \omega_j) = f_{\rm bin}^{t_{\rm end}}.
\end{equation}

Here, \( \pi_{i, t_0}^{(k)} \) and \( \pi_{i, t_\mathrm{end}}^{(k)} \) denote individual binary systems sampled at times \( t_0 \) and \( t_{\rm end} \), respectively, with \( N_{t_0} \) and \( N_{t_\mathrm{end}} \) representing the number of binaries in the corresponding snapshots. This approach avoids the need for explicit analytical forms of the BDFs, thereby reducing the complexity of dynamical effect modelling.

\section{Results and discussion}
\label{sec:results}

To analyse the simulation results, we first define the cluster boundary. In the presence of a point-mass galactic tidal field, the tidal radius serves as a natural choice—it marks the region where the galactic potential overtakes the cluster’s own potential. However, the inclusion of gas as part of the cluster’s gravitational potential significantly increases the tidal radius compared to the gas-free case, covering a large volume where the stellar density is very low (less than $1~\mathrm{M_\odot}~\mathrm{pc}^{-3}$). A system in this space has only very small probability to interact with other systems and is therefore very dynamical inactive. To define a more meaningful boundary, we instead use local stellar density as the criterion. 

Following \cite{1985ApJ...298...80C}, we define the local density for each star as the mass density within a sphere that includes its six nearest neighbours. Stars with local densities below $1~\mathrm{M_\odot}~\mathrm{pc}^{-3}$ are excluded in the calculation of the cluster's centre and the boundary of the cluster . The cluster centre is then computed as the local-mass-density-weighted average position of the remaining stars, and the cut-off radius $R_{\rm cut}$ is defined as the distance from this centre to the furthest remaining star. This approach assumes that, under isotropy, the local density reflects the global density profile. In our simulations, $R_{\rm cut}$ typically lies between 2 and 3 pc. It is determined from the final snapshot (at 0.6 Myr), and the same value is used for earlier snapshots, with the cluster centre recalculated accordingly. During the evolution, stars are ejected from the cluster due to close encounters and the cluster expands slightly (see section \ref{sec:timescales} below). Thus $R_{\rm cut}$ from the final snapshot is large enough to cover all dynamical active systems during the calculation. In the analysis of binary populations, we only consider stars within $R_{\rm cut}$. At 0 Myr i.e. the initial time, there are 30 to 50 stars out of $R_{\rm cut}$ for each sample. However this will not affect the results, because we introduce no correlation between position and binary parameters in the initial setup. The half-mass radius and related quantities are determined using all stars, including those located beyond $R_{\rm cut}$, with respect to the updated cluster centre.

Binary systems are identified following the method in \citet{1995MNRAS.277.1491K,1995MNRAS.277.1522K}, where we compute the binding energy between each star and its nearest neighbour. If the binding energy is negative, the pair is considered a binary. To detect higher-order multiples, we treat identified binaries as single particles (using their c.m. properties) and repeat the procedure to find triple and quadruple systems. In this study, we include up to quadruples, but such systems are rare (no more than five per snapshot) and are excluded from further analysis. In the following analysis, a ‘system’ denotes either a single star or a binary.

In this work, we do not aggregate the simulation outputs into a single combined dataset. Our focus is on the evolution of binary populations within individual star clusters, where local conditions and individual cluster evolution can lead to significant variation across realizations. Combining all models into one sample would smooth out these differences, making it difficult to isolate the physical mechanisms responsible for binary evolution on a cluster-by-cluster basis. Such aggregation risks introducing biases or masking dynamical signatures that are only apparent when each cluster is treated as an independent system. To preserve the individuality of each model and accurately capture robust dynamical trends, we analyse all realizations separately while identifying parameters that exhibit consistent behaviour across the ensemble. We begin by visually inspecting the distribution of various parameters across individual realizations, selecting those that show consistent behaviour. We then apply different binning methods to test whether the trends persist. For each candidate parameter, we define a dynamical operator and apply importance sampling. Only parameters with consistently behaving operators across all samples are retained. Those not shown are either strongly correlated with others or too noisy to be modelled reliably.

\subsection{Timescale relevant to binary evolution}
\label{sec:timescales}
Our model evolves the cluster only up to 0.6~Myr. A natural question arises: is this short timespan sufficient for significant changes in the BP?  As demonstrated by \citet{2011MNRAS.417.1684M}, binary populations can evolve substantially on timescales comparable to the crossing time $t_{\rm cr}$, defined as the ratio of cluster diameter to the typical stellar velocity.

We adopt the half-mass radius $R_{\rm h,m}$ as the characteristic size and the root mean square of a single velocity component, i.e. one-dimensional velocity dispersion, $\sigma_{\rm ecl}$, of systems within that radius as the velocity scale. The crossing time is thus:
\begin{equation}
\label{eq:tcrdef}
	t_{\rm cr} = \frac{2R_{\rm h,m}}{\sigma_{\rm ecl}}.
\end{equation}
In our analysis, $t_{\rm cr}$ is calculated using the initial values of $R_{\rm h,m}$ and $\sigma_{\rm ecl}$. 
In our model, the gas potential increases the velocity dispersion, so we compute $t_{\rm cr}$ directly from Equation~\eqref{eq:tcrdef} with the velocity dispersion being calculated using the centre-of-mass velocities of systems within the half-mass radius. We also show here the half-mass two-body relaxation time $t_{\rm rh}$. The definition follows from e.g.  \citet{1997A&ARv...8....1M},
\begin{equation}
	\frac{t_{\rm rh}}{t_{\rm cr}} \approx 0.138\left(\frac{R_{\rm h, m}}{2R_{\rm vir}}\right)^{3/2}\frac{N}{\operatorname{ln}\Lambda},
\end{equation}
where $R_{\rm vir}$ is the virial radius of the cluster, in our case (Plummer sphere) $R_{\rm h, m} \approx 0.77 R_{\rm vir}$, $N$ is the number of systems within $R_{\rm h, m}$ and $\operatorname{ln}\Lambda$ is the Coulomb logarithm. We use $\Lambda = 0.02N$ based on the measurement of \cite{1996MNRAS.279.1037G}. 

\begin{figure}
	\centering
	\resizebox{\hsize}{!}{\includegraphics{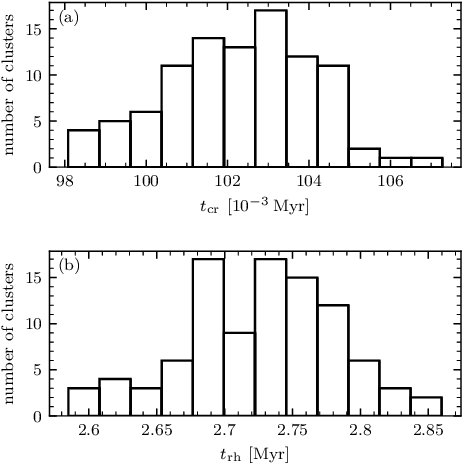}}
    \caption{Histogram of the crossing time, $t_{\rm cr}$ (panel a) and the half-mass relaxation time $t_{\rm rh}$ (panel b), for all model clusters in the sample. }
    \label{fig:tcr0_dist_S0}
\end{figure}

Figure~\ref{fig:tcr0_dist_S0}(a) shows the distribution of $t_{\rm cr}$ across all samples. Most clusters have $t_{\rm cr} \approx 0.1$ Myr. Over 0.6~Myr, the system undergoes approximately 5 to 6 crossing times. As discussed in Section~\ref{sec:initial_setup}, our setup introduces minor sample-to-sample variations, which are reflected in the distribution of $t_{\rm cr}$. The histogram exhibits a mean and median of 0.102 Myr. The skewness is $-0.07$, with a Fisher kurtosis of  -0.33. Both values indicate that the distribution of $t_{\rm cr}$ is approximately normal, exhibiting a very slight left skew and a mildly flatter shape than the standard normal curve. In \texttt{McLuster}, the half-mass radius is constrained to match the specified value; that is, $R_{\rm h, m}\equiv 0.3$ pc at the initial time. However, due to the random  nature of the velocity distribution, each realization exhibits fluctuations in the velocity dispersion. Figure~\ref{fig:tcr0_dist_S0}(b) also shows the histogram for $t_{\rm rh}$. For $t_{\rm rh}$, the number of systems within the half-mass radius (i.e., the mean system mass) introduces an additional source of dispersion. These dispersions arising from our initial setup are not considered in the present study, but their potential influence on the dynamical evolution of the cluster and the binary populations should be explored in future works.

\begin{figure*}
\sidecaption
\includegraphics[width=12cm]{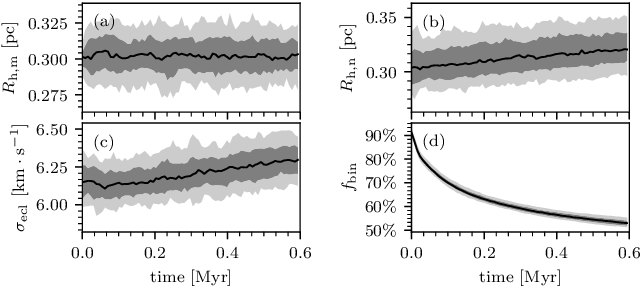}
\caption{Time evolution of global cluster properties: (a) half-mass radius $R_{\rm h,m}$; (b) half-number radius $R_{\rm h,n}$; (c) (one-dimensional) velocity dispersion within $R_{\rm h,m}$, $\sigma_{\rm cl}$; (d) binary fraction. Black lines show the median; dark and light grey areas mark the 16th–84th (68\%) and 2.5th–97.5th (95\%) percentiles, respectively.}
\label{fig:rhm_rhn_sigma_fbin_S0}
\end{figure*}

Figure~\ref{fig:rhm_rhn_sigma_fbin_S0} shows the time evolution of the half-mass radius, half-number radius, (one-dimensional) velocity dispersion, and binary fraction. The half-mass radius remains nearly constant, while the half-number radius increases by about 10\%, indicating an evolution towards mass segregation. The velocity dispersion increases by 5\%, further suggesting that the cluster structure remains largely unchanged over this timespan. However, the binary fraction decreases significantly—dropping to around 50\%—indicating that the binary population evolves much faster than the global cluster structure, because the relaxation time is $\approx 2.7$ Myr.

\subsection{Dynamical evolution of binding energy and orbital period}
\label{sec:omega_Eb_P}
The binding energy is widely used to characterize dynamical processing in star clusters. For a binary, the ratio between its binding energy and the mean kinetic energy of stars in the environment provides a natural measure of its hardness. In this section, we model the dynamical operators for both binding energy and the orbital period.

The absolute value of binding energy of a binary is defined as 
\begin{equation}
\label{eq:Ebdef}
	E_{\rm b} = G\frac{m_{\rm p}m_{\rm s}}{2a}.
\end{equation}
\citet{2011MNRAS.417.1684M} introduced a dynamical operator for $E_{\rm b}$ denoted $\Omega_{E_{\rm b}}$, parameterized by a sigmoid-like function in log-space:
\begin{equation}
	\label{eq:OmegaEb_Marks}
	\begin{aligned}
	\Omega_{E_{\rm b}}(E_{\rm b}) = \frac{\mathcal{A}^\prime_{\rm E}}{1+\exp\left[ \mathcal{S}_{\rm E}^\prime \left( \log_{10} E_{\rm b} - \mathcal{E}^\prime_{\rm cut} \right) \right]}  - \frac{\mathcal{A}^\prime_{\rm E}}{2}\\
	\text{for } \log_{10} E_{\rm b} > \mathcal{E}^\prime_{\rm cut}
\end{aligned},
\end{equation}
with $\Omega_{E_{\rm b}}(E_{\rm b}) \equiv 0$ for $\log_{10} E_{\rm b} \leq \mathcal{E}^\prime_{\rm cut}$.
Here, $\mathcal{A}^\prime_{\rm E}$ quantifies the total variation of the operator, $\mathcal{S}_{\rm E}^\prime$ sets the steepness of the transition, and $\mathcal{E}^\prime_{\rm cut}$ defines a sharp cutoff below which the operator vanishes. In this work, we adopt a similar functional form but generalize it to allow smoother transitions and a finite asymptote at low binding energies:
\begin{equation}
\label{eq:OmegaEb}
\Omega_{E_{\rm b}}(E_{\rm b}) = \frac{\mathcal{A}_{\rm E}}{1+\exp\left[ \mathcal{S}_{\rm E} \left( \log_{10} E_{\rm b} - \mathcal{E}_{\rm cut} \right) \right]} + \mathcal{B}_{\rm E}.
\end{equation}
Here, $\mathcal{A}_{\rm E}$ sets the overall amplitude, while $\mathcal{S}_{\rm E}$ and $\mathcal{E}_{\rm cut}$ control the slope and location of the transition. Unlike the formulation of \citet{2011MNRAS.417.1684M}, we introduce an additional parameter, $\mathcal{B}_{\rm E}$, which allows the operator to asymptotically approach a non-zero value at low $E_{\rm b}$. This modification removes the sharp cutoff and accounts for the continuous presence of wide (soft) binaries, which are short-lived yet persist in quasi-equilibrium due to their continual replenishment through dynamical interactions.

\begin{figure}
    \centering
    \resizebox{\hsize}{!}{\includegraphics{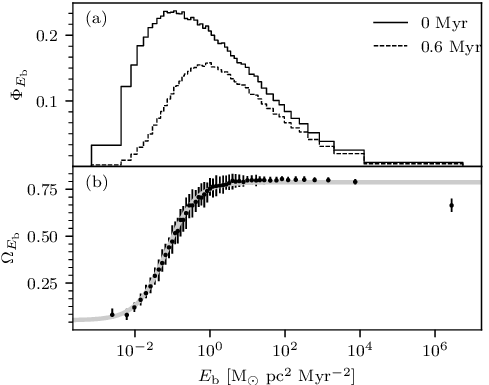}}
     \caption{(a) BDF of the absolute value of binding energy $E_{\rm b}$ at 0 Myr (solid) and 0.6 Myr (dashed). For all samples the same binning is used; curves show the median across all realizations.
     	(b) Dynamical operator $\Omega_{E_{\rm b}}$ as defined from the $N$-body data. Black dots indicate the median values, with error bars marking the 16th–84th percentiles. The solid grey line shows the median fit of Equation~\ref{eq:OmegaEb}.}
    \label{fig:Eb_omega_S0}
\end{figure}

Figure~\ref{fig:Eb_omega_S0}(b) shows the empirical (i.e. as measured from the $N$-body output data) dynamical operator $\Omega_{E_{\rm b}}$. At $E_{\rm b} \gtrsim 10^3~\mathrm{M_\odot pc^2 Myr^{-2}}$, the number of binaries becomes very small (see Figure \ref{fig:Eb_omega_S0}a), and the inferred trends in this high-energy regime are unreliable. Such strongly bound, i.e. hard, systems are either extremely compact (short-period) or contain very massive components; in both cases, stellar evolution dominates over dynamical interactions. We therefore restrict the fitting of Equation~\eqref{eq:OmegaEb} to the range $E_{\rm b}\in[0, 10^3]~\mathrm{M_\odot pc^2 Myr^{-2}}$.
For each of the 100 realizations, parameter estimation is carried out independently. To illustrate the results clearly, we show only the fit obtained from the median parameter values (solid grey line in Figure \ref{fig:Eb_omega_S0}b). The full set of fitting results is provided in Appendix\ref{sec:corner_app}. For comparison with the formulation of \citet{2011MNRAS.417.1684M}, we also perform a regression using Equation~\eqref{eq:OmegaEb_Marks}; the results and discussion are presented in Appendix~\ref{sec:marks_app}.

We note that different combinations of $\mathcal{E}_{\rm cut}$ and $\mathcal{S}_{\rm E}$ may yield similar objective function values. This degeneracy arises because a larger $\lvert \mathcal{S}_{\rm E} \rvert$ can compensate for a smaller $\mathcal{E}_{\rm cut}$. In practice, however, the new model (Equation~\ref{eq:OmegaEb}) exhibits greater robustness, since both the minimum and maximum values of $\Omega_{E_{\rm b}}$ are explicitly constrained. We further find that performing an initial $\chi^2$ fit to the histogram data, and using its outcome as a starting point for direct importance estimation, leads to more stable parameter recovery and reduces the risk of convergence to local minima.

\begin{figure}
	\centering
	\resizebox{\hsize}{!}{\includegraphics{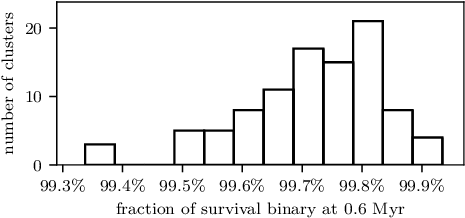}}
	\caption{Histogram of the survival fraction of binaries at 0.6 Myr for all model clusters. The survival fraction is defined as the number of binaries present at both 0 and 0.6 Myr, divided by the total number of binaries. }
	\label{fig:suvival_S0}
\end{figure}

As shown by \citet{1975MNRAS.173..729H} and \citet{1975AJ.....80..809H}, three-body encounters drive a systematic evolution of binary hardness, encapsulated in the Heggie–Hills law: hard binaries become harder, soft binaries become softer. The classification into hard or soft is inherently relative. In a single encounter, a binary is considered soft if its binding energy is smaller than the kinetic energy of the incoming single star; in this case, the single star is slowed down and the binary’s binding energy decreases in magnitude. Conversely, a binary is hard if its binding energy exceeds that kinetic energy; after the interaction, the single star is accelerated and the binary becomes more tightly bound.
In stellar clusters, this criterion is thus usually applied with respect to the average kinetic energy of single stars, such that the hard–soft boundary depends on the cluster velocity dispersion and evolves over time. Hard binaries act as an energy source, heating the cluster through single–binary interactions that accelerate stars. By contrast, soft binaries do not provide a sustained cooling effect \citep[see e.g.][]{1999NewA....4..495K}: after only a few encounters—typically within a crossing time—they are disrupted.
Taking the Heggie–Hills law into account, the dynamical operator effectively captures the net outcome of hardening, softening, mergers, disruptions, binary formation, and escape. To qualitatively assess the relative importance of these processes, we now examine each process separately:

\paragraph{Formation}  Each binary is assigned a unique identifier (as implemented in the \texttt{PeTar} analysis tool). A binary present at both 0 and 0.6 Myr is classified as a surviving binary. 
Figure~\ref{fig:suvival_S0} shows the survival fraction at 0.6 Myr, defined as the number of surviving systems divided by the total number of binaries. The result indicates that more than 99.6\% of binaries are survivors, implying that binary formation during this period is negligible, as also shown by \citet{2001ApJ...555..945K}. More complex processes, such as component exchange (where one member of a binary is replaced during an encounter) or disruption followed by subsequent re-formation, are subsumed under the formation channel in this work. These events are difficult to track in the simulations because resolving them would require output time intervals much shorter than the typical orbital periods of binaries. However, since fewer than 0.7\% of binaries at 0.6~Myr are non-surviving systems, their impact on the overall results is negligible.

\paragraph{Merging} Mergers are treated by the stellar evolution routines in \texttt{PeTar} (SSE/BSE). From the log files we find, across all model clusters, a maximum of 25 merger events and an average of about 15 per simulation. Mergers occur only in very hard binaries with extremely large binding energies. These events account for the decline of $\Omega_{E_{\rm b}}$ in the range $10^4$–$10^6~\mathrm{M_\odot pc^2 Myr^{-2}}$. Given that the maximum number of mergers (25) corresponds to only $\approx$0.6\% of the initial binary population (over 4000 systems), the overall impact of mergers is negligible. The merges are, however, of astrophysical interest as they lead to post-merger stars that do not fit for single star evolution synthesis and are likely to also show abnormal circumstellar material and magnetic activity \citep[e.g.][]{2025Galax..13...46K,2024A&A...689A.234D} 

\begin{figure}
	\centering
	\resizebox{\hsize}{!}{\includegraphics{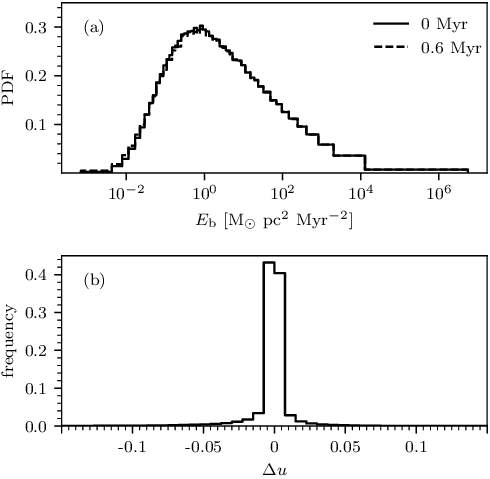}}
	\caption{(a) Empirical PDFs of the absolute value of the binding energy $E_{\rm b}$ for surviving binaries at 0 Myr (solid) and 0.6 Myr (dashed). The same binning is applied to all models, and the plotted histograms show, for each bin, the median value across all clusters. The two distributions are nearly indistinguishable.
		(b) Histogram of quantile shifts (Equation~\ref{eq:quantileshift}, see text for definition), where the bar heights represent the median values across all clusters.}
	\label{fig:Eb_su_S0}
\end{figure}

\paragraph{Hardening/Softening}
Since more than $99.3\%$ of binaries at 0.6 Myr are survivors, the final BDF is shaped almost entirely by the survivor population at 0.6 Myr. In Figure~\ref{fig:Eb_su_S0}(a) we compare the initial and evolved PDFs of $E_{\rm b}$ for the same set of surviving binaries; the two distributions are indistinguishable. A two-sample Kolmogorov–Smirnov test performed for each model cluster yields a minimum $p$-value of $0.96$, indicating no statistically significant difference.
To further examine whether substantial but mutually cancelling hardening and softening could nevertheless occur at the \emph{individual} level, we introduce a per-binary “quantile shift.” For each cluster, we first construct a single empirical cumulative distribution function (ECDF), $F(E_{\rm b})$, from the survivor sample. Given a binary with binding energy $E_{\rm b}^\prime$, its ECDF value $F(E_{\rm b}^\prime)$ corresponds to the fraction of systems with binding energy less than or equal to $E_{\rm b}^\prime$. By construction, $F(E_{\rm b})$ lies in $[0,1]$ and represents the percentile rank of that binary within the overall distribution. We then define the quantile shift between 0 Myr and 0.6 Myr as
\begin{equation}
	\label{eq:quantileshift}
	\Delta u \equiv F\left( E_{\rm b}^{0.6~{\rm Myr}} \right) - F\left( E_{\rm b}^{0~{\rm Myr}} \right),
\end{equation}
so that $\Delta u>0$ denotes hardening (a binary moves upward in the distribution) and $\Delta u<0$ denotes softening. This quantile-based measure is insensitive to the absolute scale of $E_{\rm b}$ and remains robust across its wide dynamic range.
As shown in Figure~\ref{fig:Eb_su_S0}(b), the distribution of $\Delta u$ is narrowly peaked around zero, with the overwhelming majority of binaries showing $|\Delta u|<0.05$. If strong but balanced hardening and softening were present, we would expect a broad, symmetric spread of $\Delta u$ values (many binaries shifting significantly upward and downward, even if the net PDF appears unchanged). The absence of such a spread rules out the possibility that substantial but compensating changes are hidden in the stable global PDF. We therefore conclude that, over 0–0.6 Myr, hardening and softening contribute negligibly to the evolution of the BDF.

To  directly and clearly visualize the hardening/softening evolution, Figure~\ref{fig:Eb_0_t_scatter} shows a two-dimensional density map of the absolute value of the binding energy for surviving binaries at 0~Myr and 0.6~Myr. Most surviving binaries maintain nearly unchanged binding energies, suggesting that dynamical evolution has only a minor impact on the overall orbital energy distribution. In contrast, binaries with relatively low binding energies exhibit more pronounced dynamical evolution, with both hardening and softening occurring more frequently than in tightly bound systems. The diagram further indicates that softening events are more common than hardening in our simulations. We expect that decreasing $M_{\rm ecl}$ would favour hardening over softening, since the lower velocity dispersion in such environments lets more binaries to become hard systems.

\begin{figure}
	\centering
	\resizebox{\hsize}{!}{\includegraphics{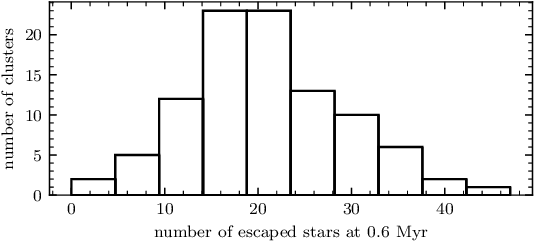}}
	\caption{Histogram for number of escaped stars at 0.6 Myr for all model clusters. The number is calculated as the difference between the number of stars within $R_{\rm cut}$ at the $t=0$ Myr and $t=0.6$ Myr.}
	\label{fig:N_esc_S0}
\end{figure}

\begin{figure}
	\centering
	\resizebox{\hsize}{!}{\includegraphics[width=\linewidth]{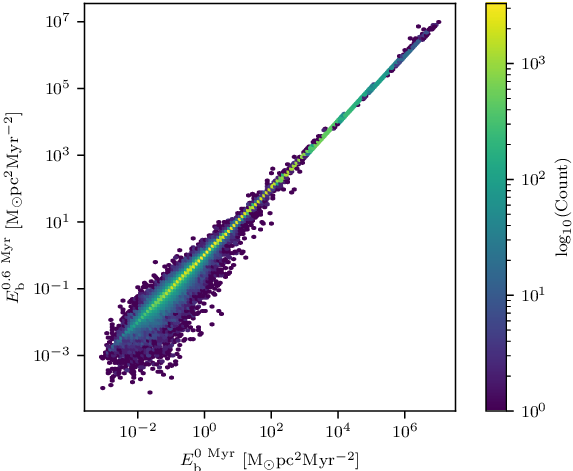}}
	\caption{Two-dimensional density map of the absolute value of the binding energy for surviving binaries at 0~Myr and 0.6~Myr. Colours indicate the number of binaries on a logarithmic scale. The main axes use an equal aspect ratio (identical scaling on both axes).}
	\label{fig:Eb_0_t_scatter}
\end{figure}

\paragraph{Escape} Because the gas potential is included as part of the cluster, the overall gravitational potential is deeper than in the gas-free case modelled by \cite{2011MNRAS.417.1684M}, and only a small number of escapers are expected. We define escapers as stars located beyond $R_{\rm cut}$ at 0.6 Myr. Across all model clusters, the number of escapers never exceeds 47 (Figure~\ref{fig:N_esc_S0}). Compared to the total binary population of several thousand, this number is negligible. We therefore conclude that stellar escape does not play a significant role in shaping the evolution of the binary distribution at this stage.

\paragraph{Disruption} From the preceding discussion, we find that formation, merging, escape, and hardening/softening all have negligible impact. Consequently, binary disruption emerges as the dominant dynamical effect within the first 0.6 Myr. Disruption efficiently removes soft binaries from the population and therefore shapes the low-binding-energy tail of the distribution.
It is important to note, however, that our survival-probability–based approach imposes intrinsic limitations. By construction, it can only track binaries that remain bound and thus cannot account for the appearance of new wide systems. In other words, this method constrains the outcome within a restricted binding-energy range, but cannot reproduce the continuous replenishment of wide binaries through dynamical encounters. As a result, the present ansatz is best interpreted as valid only between lower and upper limits in binding energy, outside of which additional physics—particularly formation and re-formation of wide binaries—must be invoked.

In summary, among the various dynamical processes considered—formation, merging, escape, and hardening/softening—only disruption has a significant impact. Accordingly, $\Omega_{E_{\rm b}}(E_{\rm b})$ can be interpreted as the survival probability of binaries at a given binding energy, consistent with the formulation of \citet{2011MNRAS.417.1684M}.

\begin{figure}
	\centering
	\resizebox{\hsize}{!}{\includegraphics{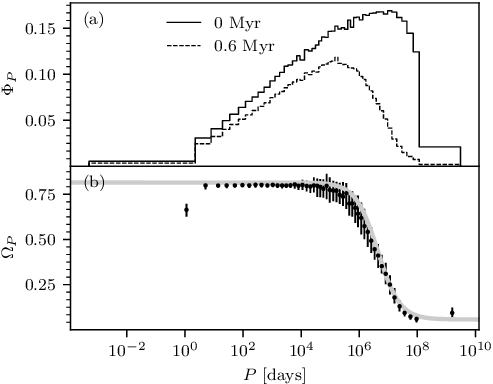}}
	\caption{Empirical BDF and dynamical operator for the orbital period $P$. The panel layout is analogous to Figure~\ref{fig:Eb_omega_S0}. A clear cut-off is seen around $10^7$ days; model fitting is restricted to $P > 2$ days, where dynamical effects dominate over stellar evolution. A slight deviation is visible around $E_{\rm b}\approx10^{5}$–$10^{6}$, reflecting a limitation of our model in accurately capturing the evolution of binaries in this period range.}
	\label{fig:P_omega_S0}
\end{figure}

Another key observable is the orbital period $P$, which can be directly constrained from optical observations. Figure~\ref{fig:P_omega_S0} shows the empirical BDF and dynamical operator for $P$, where a cut-off is evident around $10^7$ days. We adopt the same functional form as in Equation~\eqref{eq:OmegaEb} for the dynamical operator, but with distinct parameters: $\mathcal{A}_{\rm P}$, $\mathcal{S}_{\rm P}$, $\mathcal{P}_{\rm cut}$, and $\mathcal{B}_{\rm P}$. For the fitting we restrict the sample to $P > 2$ days, since systems with shorter periods are extremely tight and their evolution is dominated by stellar evolution rather than dynamics.
In addition, we compare the initial and evolved PDFs of $P$ for surviving binaries, as well as their quantile shift (see Appendix~\ref{app:ppdf}). The results mirror those obtained for the binding energy: the distributions remain essentially unchanged, and disruption is the only significant dynamical effect on the orbital-period distribution. Thus the period operator essentially encodes the survival probability in $P$, analogous to the binding-energy case.

Both the positive offsets $\mathcal{B}_{\rm E}$ and $\mathcal{B}_{\rm P}$ in the fitted dynamical operators imply that soft binaries are not disrupted instantaneously. Instead, they can persist transiently, and in some cases escape intact if they form near the cluster periphery. Such escaping wide binaries may naturally contribute to the very wide binary population observed in the Galactic field.

\subsection{Mass-ratio evolution}
\label{sec:mass_ratio}
Unlike binding energy and orbital period, where the survival probability offers a relatively direct diagnostic of dynamical processing, the evolution of the mass-ratio distribution reflects a more complex interplay between internal stellar evolution, dynamical encounters, and pre-main-sequence eigenevolution.

\begin{figure}
  \centering
  \resizebox{\hsize}{!}{\includegraphics{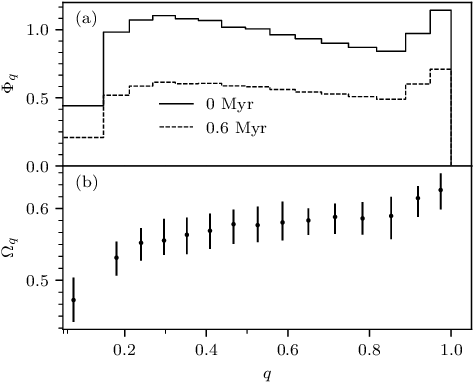}}
  \caption{(a) BDF of the mass ratio $q$ at $t=0$ Myr (solid) and $t=0.6$ Myr (dashed). For each sample the same binning is used; curves show the median across all clusters.
  	(b) Empirical dynamical operator $\Omega_q$. Black dots indicate the median values, with error bars marking the 16th–84th percentile range.}
  \label{fig:q_omega_S0}
\end{figure}

\begin{figure*}
	\sidecaption
	\includegraphics[width=12cm]{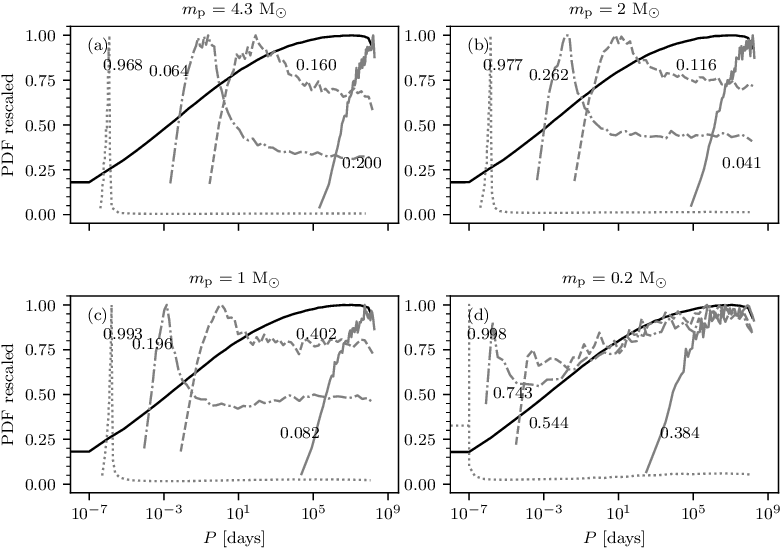}
	\caption{Initial (i.e. after pre-main-sequence eigenevolution) period distributions for binaries with different mass ratios and primary masses. Each panel corresponds to a different primary mass (as indicated in the title). The black solid line shows the overall period distribution for that primary mass, while grey lines (with different styles) indicate distributions in individual $q$ bins. The label next to each gray line denotes the left edge of the corresponding $q$ bin; all bins have a fixed width of 0.001. All distributions are normalized to their peak values to facilitate visual comparison across different scales. Note that the extremely short orbital period appearing here ($\approx 10^{-7}$ days, i.e., on the order of milliseconds) is unphysical. Such systems are rarely generated and would merge almost immediately after the start of the simulation.}
	\label{fig:mqP_S0}
\end{figure*}

Figure~\ref{fig:q_omega_S0} shows the BDF and empirical dynamical operator for $q$. Owing to the substantial scatter among different realizations, we do not attempt a parametric fit to the operator. Nevertheless, a clear trend emerges: binaries with low mass ratios ($q \lesssim 0.2$) are strongly suppressed, whereas those with nearly equal masses ($q \approx 1$) are preferentially retained, with the operator reaching values of $\approx 0.6$. In the intermediate regime, the operator remains approximately constant at $\approx 0.55$.
This behaviour can be understood in the framework of pre-main-sequence eigenevolution, in which the birth mass ratio, $q_{\rm bir}$, of low-mass binaries is modified to the initial ratio $q_{\rm ini}$ according to  \citep[as described by][]{1995MNRAS.277.1491K,1995MNRAS.277.1522K}
\begin{equation}
	q_{\rm ini} =
	\begin{cases}
		q_{\rm bir} + \rho(1-q_{\rm bir}) & \rho \leq 1 \\
		1 & \rho > 1
	\end{cases},
\end{equation}
where
\begin{equation}
	\rho = \left(\frac{\lambda \mathrm{R_\odot}}{r_{\rm peri}}\right)^{\chi},
\end{equation}
with $\lambda = 28$, $\chi = 3/4$, and $r_{\rm peri}$ the pericentre distance in units of solar radii. The corresponding modification of eccentricity is given by
\begin{equation}
	\operatorname{ln} e_{\rm ini} = -\rho + \operatorname{ln} e_{\rm bir}.
\end{equation}
These expressions indicate that long-period low-mass binaries experience minimal change in eccentricity due to pre-main-sequence eigenevolution, while the mass ratio is strongly altered for close systems. We generate a large number of such systems with different primary masses ($m_{\rm p} = 4.3$, 2.0, 1.0, 0.21~$\mathrm{M_\odot}$) to illustrate how the initial period distribution depends on $q$. As shown in Figure~\ref{fig:mqP_S0}, binaries with $q \approx 1$ concentrate strongly at very short periods, where they may rapidly merge. Conversely, systems with very low $q$ are limited to long periods (from thousands of years to about 1 Myr), which fall within the range of strong dynamical processing (cf. Figure~\ref{fig:P_omega_S0}). Hence, we expect dynamical effects to preferentially modify low-$q$ binaries. 
Furthermore, we find that as the primary mass decreases, the separation between period distributions corresponding to intermediate $q$ values narrows, while the difference at very low $q$ persists. This can be attributed to the limited range of physically possible $q$ values for low-mass primaries due to the lower mass cut-off in the IMF. 
Overall, these results suggest that the apparent role of the mass ratio arises mainly from its correlation with binding energy established during pre-main-sequence eigenevolution, rather than from any intrinsic dynamical effect of $q$ itself.

\subsection{Mass dependence of the binary fraction}
\label{sec:mass_dependence_fbin}
Observational studies \citep[e.g.][]{2003MNRAS.346..354K,2014MNRAS.437.1216D,2023ASPC..534..275O} have consistently shown that the observed binary fraction of Galactic field stars depends strongly on the primary stellar mass. To quantify this relation, we define
\begin{equation}
	f_{\rm bin}(m_{\rm p}) = \frac{N_{\rm bin}(m_{\rm p})}{N_{\rm bin}(m_{\rm p})+ N_{\rm single}(m_{\rm p})}
\end{equation}
where $N_{\rm bin}(m_{\rm p})$ and $N_{\rm single}(m_{\rm p})$ denote the number of binaries and single stars with primary mass $m_{\rm p}$, respectively. Here we consider only binary systems, and neglect higher-order multiples (triples, quadruples, etc.). This definition is directly comparable to observations: when a sample is constructed by selecting stars within a given mass range and binaries are classified according to their primary mass, the measured binary fraction corresponds to $f_{\rm bin}(m_{\rm p})$ as defined above. In this sense, $f_{\rm bin}(m_{\rm p})$ represents the probability that a star of mass $m_{\rm p}$ has a companion of lower mass ($q \leq 1$). By construction of our initial conditions, all stars are assumed to reside in binaries, such that $f_{\rm bin}(m_{\rm p}) \equiv 1$ at $t=0$ Myr. The subsequent time evolution of $f_{\rm bin}$ therefore directly reflects the combined influence of dynamical processes, and provides a natural basis for comparison with observational constraints.

\begin{figure}
    \centering
   \resizebox{\hsize}{!}{\includegraphics{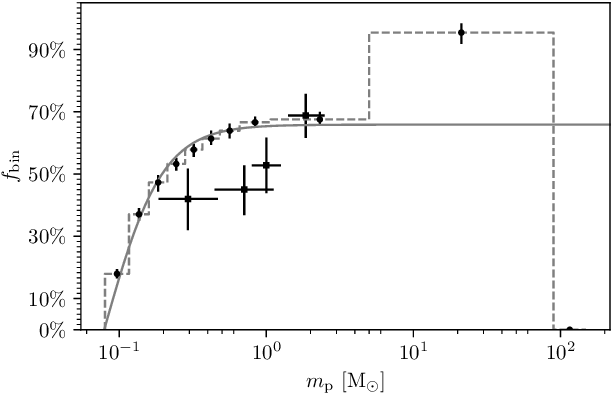}}
    \caption{Binary fraction as a function of primary mass $m_{\rm p}$. Points show the median across all samples, with error bars marking the 16th–84th percentile range. The dashed gray line indicates the binning scheme and median values. The last bin spans $89.73$–$149.17~{\rm M_\odot}$; its upper edge corresponds to the most massive stars found in our models, which form through mergers and are therefore single. The solid gray curve shows the best-fit model obtained via maximum-likelihood estimation. Observational results are over-plotted as open squares: \citet{2003MNRAS.346..354K} for M-, K- and G-type stars (left three points) and \citet{2014MNRAS.437.1216D} for A-type stars (rightmost point). Note, all observational results come from the Galactic field.}
    \label{fig:fbin_mbin_S0}
\end{figure}

Figure~\ref{fig:fbin_mbin_S0} shows the measurement of $f_{\rm bin}(m_{\rm p})$ from the model clusters  at 0.6 Myr. A clear dependence on primary mass is evident. For $m_{\rm p} > 5~\mathrm{M_\odot}$, binaries are dynamically stable and rarely disrupted, since their large absolute value of binding energies place them firmly in the hard-binary regime. As discussed in section~\ref{sec:omega_Eb_P}, such systems are only affected by mergers. This result is consistent with observations indicating that nearly all massive stars reside in binaries \citep[see e.g.,][]{2023ASPC..534..275O}. For $m_{\rm p} < 5~\mathrm{M_\odot}$, the binary fraction rises steeply with increasing mass, saturating near $1~\mathrm{M_\odot}$ and approaching an asymptotic value of $\approx 65\%$. Above $\approx 0.3~\mathrm{M_\odot}$, however, in our model results, the growth of $f_{\rm bin}$ slows considerably, indicating a turnover in the low-mass regime where the binary fraction decreases to very low values.

To characterize the trend for $m_{\rm p} < 5~\mathrm{M_\odot}$, we adopt the same sigmoid-like functional form as in Equation~\eqref{eq:OmegaEb_Marks}, but without a hard cut-off:
\begin{equation}
\label{eq:fbinmp}
f_{\rm bin}(m_{\rm p}) = \frac{\mathcal{A}_\mathrm{b}}{1+\exp\left[\mathcal{S}_\mathrm{b} \left( \log_{10} m_{\rm p} - \mathcal{M}_{\rm cut} \right)\right]} - \frac{\mathcal{A}_\mathrm{b}}{2}.
\end{equation}
Here, $\mathcal{A}_\mathrm{b}/2$ sets the saturation level of $f_{\rm bin}$ for intermediate-mass primaries, while $10^{\mathcal{M}_{\rm cut}}$ defines the characteristic mass scale below which the binary fraction declines toward zero.

We interpret $f_{\rm bin}(m_{\rm p})$ probabilistically, and model the binary indicator variable with a Bernoulli distribution. The corresponding log-likelihood function for the parameters $\mathcal{A}_\mathrm{b}$, $\mathcal{S}_\mathrm{b}$, and $\mathcal{M}_{\rm cut}$ is :
\begin{equation}
\begin{aligned}
\ln \mathcal{L}(\mathcal{A}_\mathrm{b}, \mathcal{S}_\mathrm{b}, \mathcal{M}_{\rm cut}) = \sum_i \Bigg[ y_i \cdot \ln f_{\rm bin}(m_{\rm p}^{(i)}) \\
+ (1 - y_i) \cdot \ln\left(1 - f_{\rm bin}(m_{\rm p}^{(i)}) \right) \Bigg],
\end{aligned}
\end{equation}
where $y_i = 1$ if the $i$-th system is a binary with $q \leq 1$, and $y_i = 0$ otherwise. For each sample we perform maximum-likelihood estimation restricted to $m_{\rm p} < 5~\mathrm{M_\odot}$. The solid curve in Figure~\ref{fig:fbin_mbin_S0} shows the median fit across all samples.

Comparison with the observational results for G-, K-, M-, and A-type stars in the Galactic field \citep{2003MNRAS.346..354K,2014MNRAS.437.1216D}, and shown in Figure~\ref{fig:fbin_mbin_S0}, reveals that our models predict a higher fraction of low-mass binaries. For A-type stars, by contrast, the model reproduces the observations well. The binary fraction at a given mass can be reduced in two ways: (i) by the disruption of binaries whose primaries fall in that mass range, and (ii) indirectly, through the disruption of higher-mass binaries. $f_{\rm bin}(m_{\rm p})$ is defined with respect to the primary mass; it is calculated by counting all single stars and binary systems with primary mass $m_{\rm p}$ within a given mass range and identifying the fraction of binary systems. When a binary system is disrupted, its contribution to the binary population at the primary mass $m_{\rm p}$ is removed, while the former secondary becomes a single star at its own mass bin. As a result, the number of binary systems at the secondary mass $m_{\rm s}$ is not reduced by this disruption however the number of single stars at $m_{\rm s}$ is increased by 1 and thus the binary fraction also reduced\footnote{For example, consider a binary system composed of a G-type primary and a M-type secondary. If such a system is disrupted, the binary fraction of G-type stars is reduced and simultaneously the binary fraction of M-type stars is also decreased because one more single stars is added into the M-type stars. }. If, at later evolutionary stages, disruption becomes inefficient for binaries with $m_{\rm p}\gtrsim 1~\mathrm{M_\odot}$, then the further decline of the binary fraction at lower masses must be driven primarily by the  preferential dynamical  disruption of low-mass systems i.e. channel (i). It is possible that post–gas-expulsion evolution continues to play an important role for binaries with $m_{\rm p}\lesssim 1~\mathrm{M_\odot}$, which are characterized by low binding energies and are particularly vulnerable to perturbations.
This effect is expected to differ between the Galactic field and cluster environments, potentially leading to systematic differences in the binary population properties of the two. A more complete treatment will require modelling the subsequent dynamical evolution of low-mass binaries beyond the embedded phase.

\subsection{Future works}
Future work will investigate several key factors that may influence the dynamical evolution of binary populations in embedded clusters. 

First, variations in the total cluster mass will be examined in combination with different star formation efficiencies (SFE). These two parameters jointly determine the velocity dispersion of the system, which in turn sets the critical binding energy threshold distinguishing hard and soft binaries. \citet{2011MNRAS.417.1684M} have demonstrated that the dynamical evolution of binary populations depends primarily on the initial density of the cluster. By varying the cluster's initial mass and half-mass radius, they showed that the fitting parameters describing binary processing correlate strongly with cluster density. It will be explored whether similar trends emerge in the present models. 

Secondly, hardening and softening effects are expected to be more prominent in low-mass clusters, where the velocity dispersion is relatively low. In such environments, binary encounters are less energetic and thus more likely to result in gradual energy exchange rather than immediate disruption. Consequently, binaries tend to evolve through cumulative interactions—either hardening or softening—rather than being promptly destroyed, as is more common in high-velocity environments where encounters are more violent.

In addition to the above factors, a third aspect worth attention concerns the role of high-mass binaries. The present work does not explore in detail the dynamical evolution of high-mass binary systems. Due to the shape of the IMF, such systems are intrinsically rare in the models. However, in more massive environments such as young globular clusters, the number of high-mass binaries can be significantly larger, making their dynamical processing and resulting populations more relevant, in particular for the emergence of multiple population \citep[see e.g.][]{2020MNRAS.491..440W}. Recent observational studies have highlighted the importance of high-mass binaries: for instance, a correlation between their properties and metallicity has been reported \citep{2025NatAs...9.1337S}, and pulsar observations have revealed unusually wide binary systems possibly linked to the evolution of massive binaries \citep{2025ApJS..279...51L}. These findings suggest that the formation and survival of high-mass binaries remain an open question and warrant further investigation.
\section{Summary and conclusions}
\label{sec:conclusion}
In this study, we investigated the early dynamical evolution of binary populations in embedded massive star clusters using direct $N$-body simulations with \texttt{PeTar}. By systematically comparing binary distribution functions (BDFs) at $t=0$ and $t=0.6$ Myr, we quantified the impact of stellar dynamics through empirical dynamical operators. Constructing such operators proves to be highly non-trivial; our results show that only the binding energy and orbital period provide robust dynamical diagnostics, both of which can be described by sigmoid-like operators with interpretable parameters. This highlights the necessity of direct $N$-body simulations for modelling embedded clusters, as no simple analytic substitute exists. 

Compared to the model of \citet{2011MNRAS.417.1684M}, our formulation introduces an additional offset term, enabling the operator to account for a residual population of wide binaries. This refinement is crucial to capture the persistence of wide systems, which may escape the cluster before disruption and plausibly contribute to the very wide binaries observed in the Galactic field.  At the same time, in line with \citet{2011MNRAS.417.1684M}, it has not been explicitly tested whether the binding energy alone already provides a complete description of binary evolution.

Other binary properties exhibit weaker dynamical signatures. The shape of the mass-ratio distribution shows no strong evolutionary trend, being shaped primarily by the initial conditions and pre-main-sequence eigenevolution rather than cluster dynamics. At 0.6 Myr, the binary fraction as a function of primary mass, $f_{\rm bin}(m_{\rm p})$, shows a strong mass dependence that can be modeled by a sigmoid-like function: high-mass stars ($m_{\rm p}\gtrsim5,M_\odot$) almost universally retain companions, while the fraction declines toward lower masses. This trend is consistent with observations of A-type stars, although our models predict a slightly larger fraction of  M-, L- and G-type star-binaries than in the Galactic field.

Taken together, these results suggest a coherent picture in which disruption dominates the early dynamical evolution, primarily affecting soft, low-binding-energy binaries, while the shape of the  mass ratio distribution is largely set by pre-main-sequence eigenevolution prior to dynamical interactions. Thus, the key dynamical information is effectively encoded in binding energy and orbital period, with other parameters playing a secondary role.

For modelling the binary population of an individual cluster, direct $N$-body simulations remain indispensable. On galactic scales, however, where the field population originates from the dissolution of many embedded clusters with diverse properties, population synthesis approaches in the spirit of \citet{2011MNRAS.417.1684M} remain valuable—provided that realistic cluster-scale dynamics, such as those quantified here, are incorporated. Future extensions of this model will further clarify the connection between early cluster dynamics and the present-day binary population.
\begin{acknowledgements}
PK acknowledges support through the DAAD Eastern European Exchange Programme between the University of Bonn and Charles University in Prague as well as through grant 26-21774S from the Czech Grant Agency.
\end{acknowledgements}

\bibliographystyle{aa}
\bibliography{ref}
\begin{appendix}
	\section{PDF and quantile shift distribution of Period}
	\label{app:ppdf}
\begin{figure}
		\centering
		\resizebox{\hsize}{!}{\includegraphics{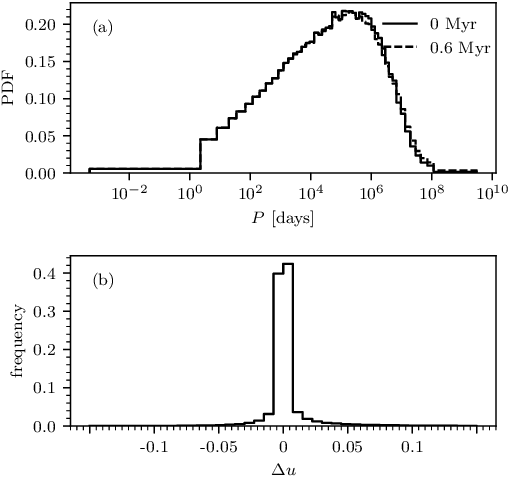}}
		\caption{(a) Empirical PDFs of the period $P$ for surviving binaries at 0 Myr (solid) and 0.6 Myr (dashed). The same binning is applied to all models, and the plotted histograms show, for each bin, the median value across all clusters. The two distributions are nearly indistinguishable.
			(b) Histogram of quantile shifts, where the bar heights represent the median values across all clusters.}
		\label{fig:P_su_S0}
	\end{figure}

	Figure~\ref{fig:P_su_S0} shows the initial and evolved PDFs, together with the quantile shifts for the periods of surviving binaries. The result is consistent with that obtained for the absolute value of binding energy: the dynamical operator can be interpreted as a disruption probability. Since orbital period and binding energy are strongly correlated, binding energy alone provides a sufficient and physically well-motivated basis for population synthesis models. The analysis of the period distribution therefore mainly serves as a cross-check of the binding-energy results.
	\section{Results on fitting parameters}
	\label{sec:corner_app}
\begin{figure}
		\centering
		\resizebox{\hsize}{!}
		{\includegraphics{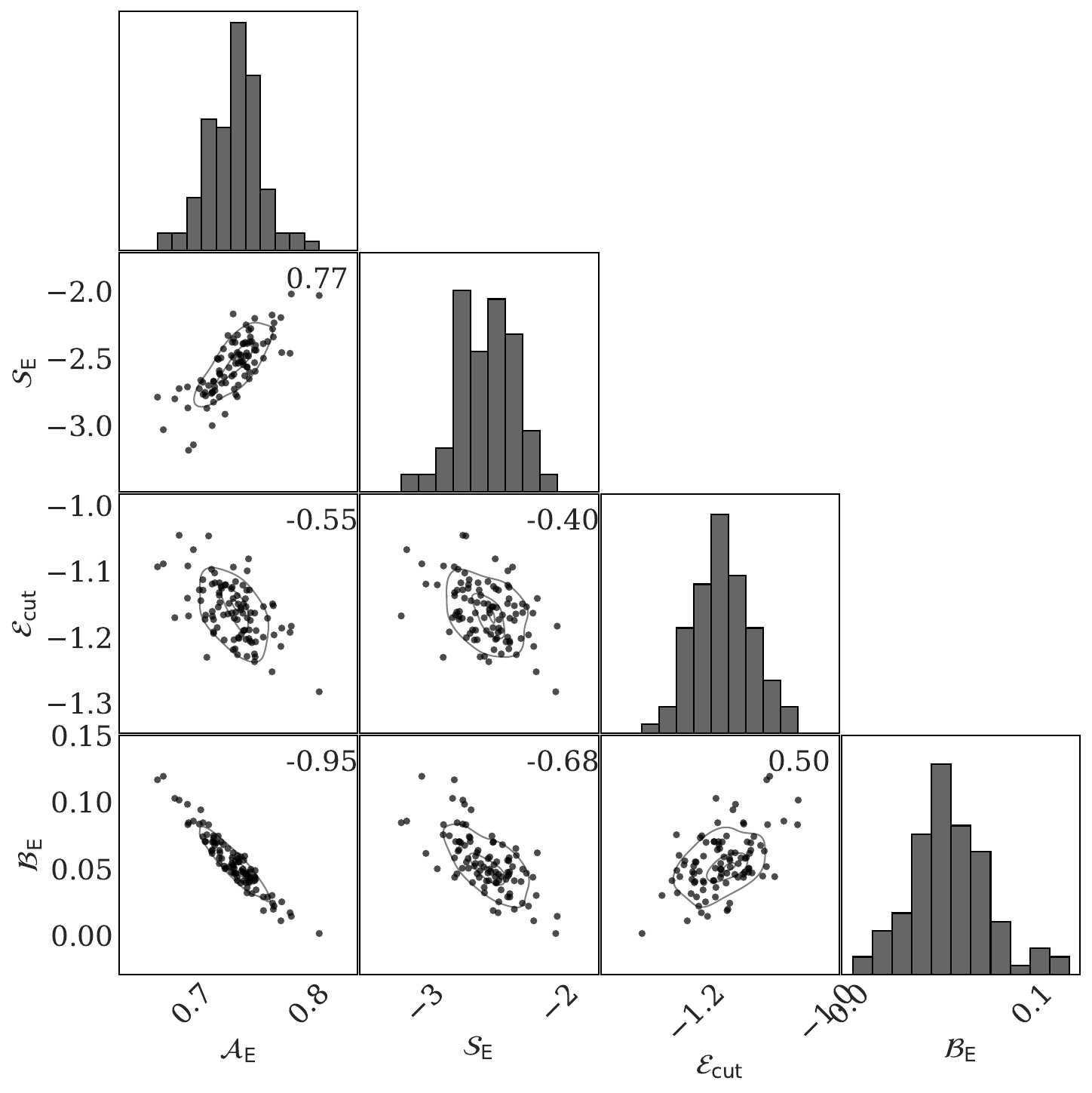}}
		\caption{Corner plots of the fitted parameters $\mathcal{A}_{\rm E}$, $\mathcal{S}_{\rm E}$, $\mathcal{E}_{\rm cut}$, and $\mathcal{B}_{\rm E}$. Contours in the scatter plots indicate the 39.3\% ($1\sigma$), 86.5\% ($2\sigma$), and 98.9\% ($3\sigma$) confidence levels. Numbers in the upper right of each panel give the Pearson correlation coefficient between the corresponding parameters.
			(a) Parameters .
			(b) Parameters $\mathcal{A}_{\rm P}$, $\mathcal{S}_{\rm P}$, $\mathcal{P}_{\rm cut}$, and $\mathcal{B}_{\rm P}$.
			Each point corresponds to one simulation, not posterior samples.}
		\label{fig:corner_Eb_S0}
	\end{figure}

\begin{figure}
		\centering
		\resizebox{\hsize}{!}
		{\includegraphics{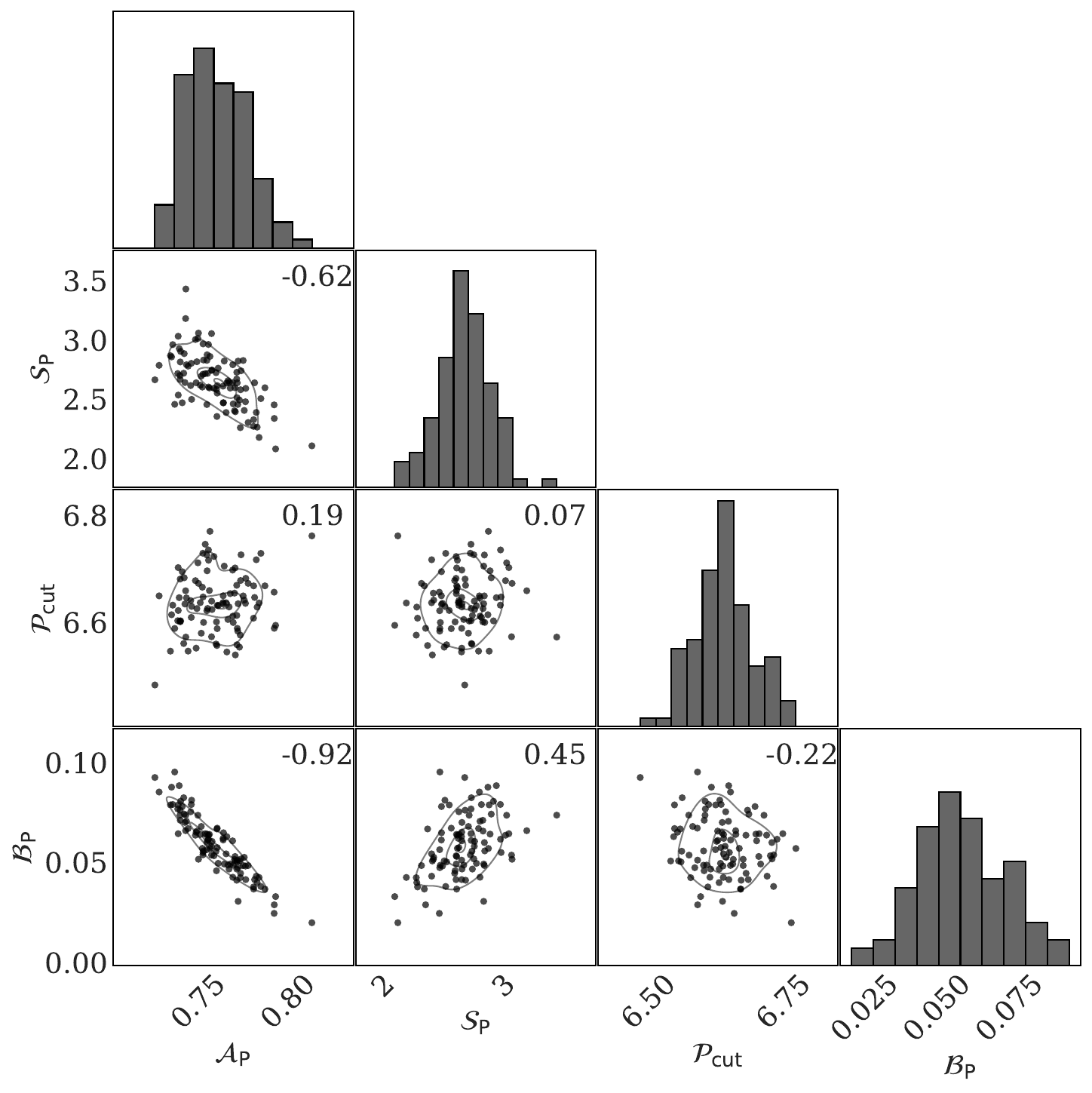}}
		\caption{Corner plots of the fitted parameters $\mathcal{A}_{\rm P}$, $\mathcal{S}_{\rm P}$, $\mathcal{P}_{\rm cut}$, and $\mathcal{B}_{\rm P}$. Plot formatting is consistent with Figure~\ref{fig:corner_Eb_S0}. }
		\label{fig:corner_P_S0}
	\end{figure}

\begin{figure}
	    \centering
	    \resizebox{\hsize}{!}{\includegraphics{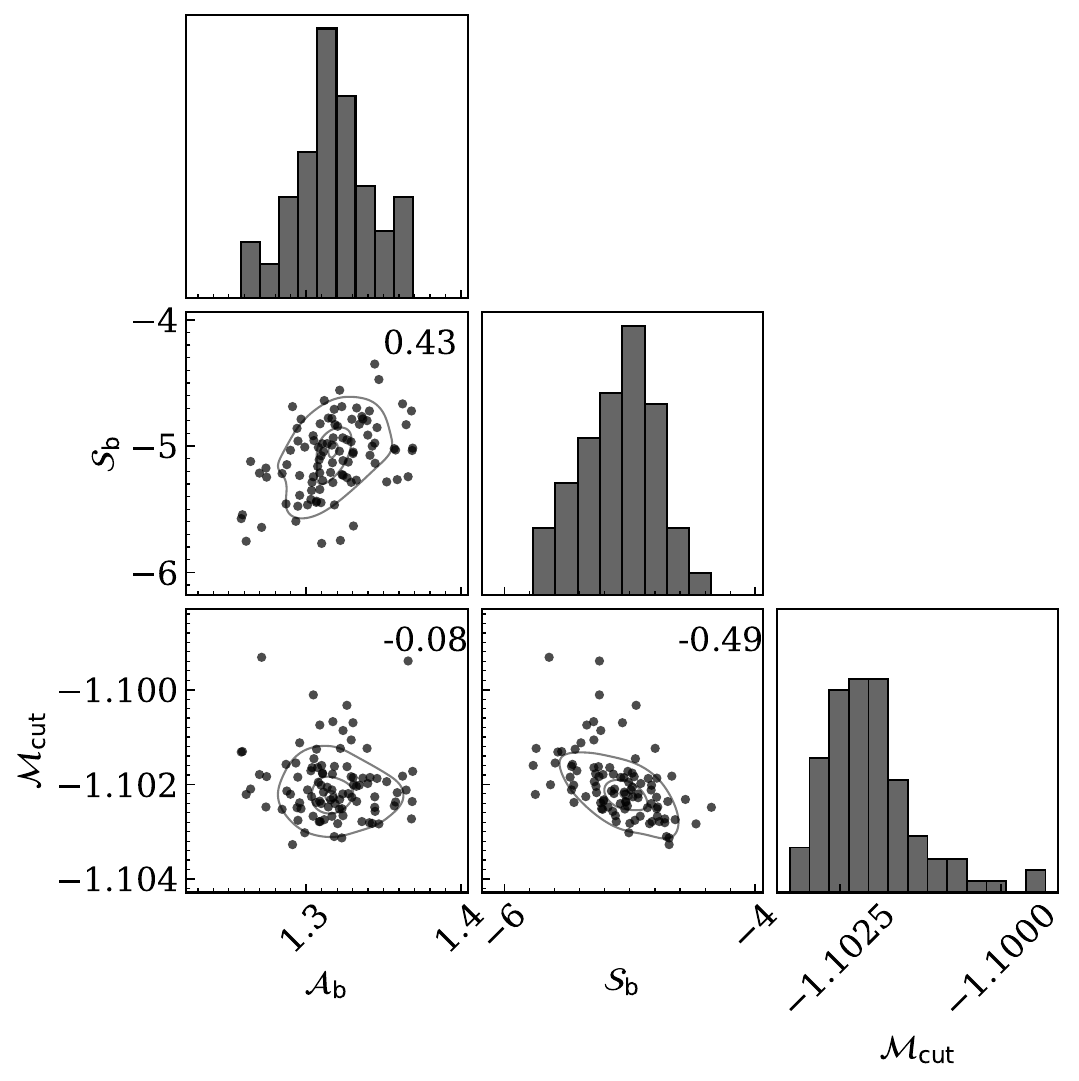}}
	    \caption{Corner plot of the fitting parameters $\mathcal{A}_{\rm b}$, $\mathcal{S}_{\rm b}$, and $\mathcal{M}_{\rm cut}$. Plot formatting is consistent with Figure~\ref{fig:corner_Eb_S0}. }
	    \label{fig:corner_fbin_mp_S0}
	\end{figure}

	Tables~\ref{tab:Eb_parameter}, \ref{tab:P_parameter} and \ref{tab:fbin_mp_parameter} list several percentiles of the fitted parameters across all models, highlighting their statistical properties. To further illustrate both the scatter among different realizations and the correlations between parameters, we visualize their distributions as corner plots in Figure~\ref{fig:corner_Eb_S0}, \ref{fig:corner_P_S0} and~\ref{fig:corner_fbin_mp_S0} .
	
	We emphasize that the corner plots shown here should not be interpreted as posterior distributions. Instead, each point represents the best-fit parameter set from a single simulation. All simulations share the same physical setup, but differ in their random initial conditions. The use of corner plots therefore serves two purposes: (i) to demonstrate how randomness in the initial cluster configuration leads to variation in the fitted parameters, and (ii) to provide an intuitive visualization of possible correlations among them.
	
	The corner plots reveal a strong anti-correlation between $\mathcal{A}_{\rm E/P}$ and $\mathcal{B}_{\rm E/P}$ (Pearson correlation coefficient $<-0.9$), suggesting that an abundance of hard binaries suppresses the survival or appearance of wide systems. A plausible explanation is that frequent encounters with hard binaries heat the cluster and disrupt wide pairs. Binary--binary encounters can directly destroy wide systems, while binary--single interactions harden the hard binaries and inject energy into the cluster. The resulting heating increases the velocity dispersion and reduces the chance for wide binaries to survive or form. By contrast, we find no significant correlations between other parameter pairs.
	
	\begin{table}
		\centering
		\caption{Fitting results for $\Omega_{E_{\rm b}}$.}
		\begin{tabular}{c c c c c c}
			\hline\hline
			Parameter  & 2.5\% & 16\% & 50\% &  84\%  &   97.5\% \\ \hline
			$\mathcal{A}_{\rm E}$  &  0.686 &  0.714 &  0.736 &  0.753 & 0.780 \\
			$\mathcal{S}_{\rm E}$  &  -3.016 & -2.756 & -2.535 & -2.360 & -2.172 \\
			$\mathcal{E}_{\rm cut}$ & -1.235 & -1.204 & -1.163 & -1.119 & -1.072 \\
			$\mathcal{B}_{\rm E}$  &   0.015 &  0.037 &  0.052 &  0.074  & 0.102 \\
			\hline
		\end{tabular}
		\tablefoot{ Parameters are defined in Eq.~\ref{eq:OmegaEb}. For all models combined, we report the 2.5\% ($-2\sigma$), 16\% ($-1\sigma$), 50\% (median), 84\% ($+1\sigma$), and 97.5\% ($+2\sigma$) percentiles of each parameter, representing their statistical distributions. These values correspond to the points shown in the corner plots (Figure~\ref{fig:corner_Eb_S0}).}
		\label{tab:Eb_parameter}
	\end{table} 
	
	\begin{table}
		\centering
		\caption{Fitting results for $\Omega_{P}$.}
		\begin{tabular}{c c c c c c}
			\hline\hline
	 Parameter  & 2.5\% & 16\% & 50\% &  84\%  &   97.5\% \\ \hline
			$\mathcal{A}_{\rm P}$ &    0.734 &  0.741 & 0.758 & 0.775 & 0.792 \\
			$\mathcal{S}_{\rm P}$  &   2.213 &  2.456 &  2.642 &  2.862 &  3.059 \\
			$\mathcal{P}_{\rm cut}$ & 6.545  &6.585 & 6.635  &6.697 &  6.742 \\
			$\mathcal{B}_{\rm P}$  &   0.030 &  0.043 &  0.057 &  0.075  & 0.088 \\
			\hline
		\end{tabular}
		\tablefoot{Same as Table~\ref{tab:Eb_parameter}, but for the fitting parameters of $\Omega_{P}$. The corresponding corner plots are shown in Figure~\ref{fig:corner_P_S0}.}
		\label{tab:P_parameter}
	\end{table} 
	\begin{table}
		\centering
		\caption{Fitting results for $f_{\rm bin}(m_{\rm p})$.}
		\begin{tabular}{c c c c c c}
			\hline\hline
			Parameter  & 2.5\% & 16\% & 50\% &  84\%  &   97.5\% \\ \hline
			$\mathcal{A}_{\rm b}$ &    1.262 &  1.295 &  1.317 & 1.342 &  1.367 \\
			$\mathcal{S}_{\rm b}$  &  -5.706 & -5.411& -5.059& -4.787 & -4.590 \\
			$\mathcal{M}_{\rm cut}$ & -1.103 & -1.103 & -1.102 & -1.101 & -1.100 \\
			\hline
		\end{tabular}
		\tablefoot{Parameters are defined in Eq~\ref{eq:fbinmp}. Same as Table~\ref{tab:Eb_parameter}, but for the fitting parameters of $f_{\rm bin}(m_{\rm p})$. The corresponding corner plots are shown in Figure~\ref{fig:corner_fbin_mp_S0}.}
		\label{tab:fbin_mp_parameter}
	\end{table} 

	\section{Comparison to \citet{2011MNRAS.417.1684M}}
	\label{sec:marks_app}
	To directly compare with previous work, we apply the dynamical operator given by Eq.~(\ref{eq:OmegaEb_Marks}) to fit the evolution of the absolute value of binding-energy distribution under our updated IBP prescription and including the gas potential. For this comparison we adopt a simple $\chi^2$ fit to the histograms, which avoids the additional complexity that a hard cut-off would introduce in a direct importance-estimation framework.
	
	To enable a more intuitive comparison, we map our fitted model parameters back to the framework of \citet{2011MNRAS.417.1684M}. In their formulation, the dynamical operator is explicitly parameterized as a function of the initial cluster density $\rho_{\rm ecl}$. By applying their equations (21)–(23), we transform each of our fitted parameters into an equivalent cluster density, which we denote as $\rho_{\rm ecl,M}$. This mapping provides a common basis for comparison: instead of contrasting operator parameters directly (which depend on our modified functional form), we can identify which effective $\rho_{\rm ecl,M}$ values in the  \citet{2011MNRAS.417.1684M} framework are dynamically equivalent to our embedded cluster models, i.e. reproduce the behaviour of our simulations. In other words, this approach allows us to directly place our results onto the $\rho_{\rm ecl}$ scale used by \citet{2011MNRAS.417.1684M}, thereby highlighting both agreements and systematic differences.
	
\begin{figure}
		\centering
		\resizebox{\hsize}{!}{
			\includegraphics{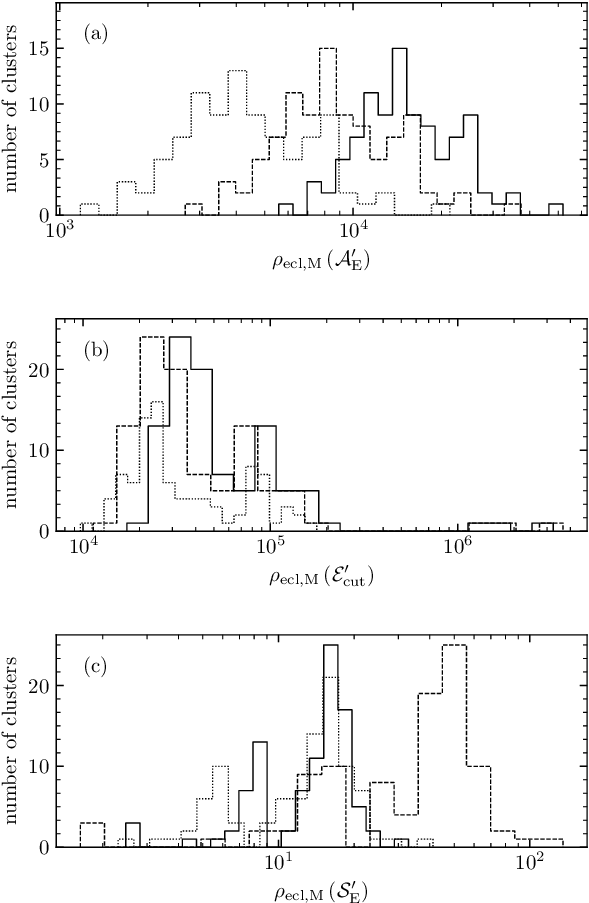}
		}
		\caption{Equivalent cluster densities $\rho_{\rm ecl,M}$, shown as histograms. Solid, dashed, and dotted lines correspond to values obtained using the 1, 3, and 5 Myr models of \citet{2011MNRAS.417.1684M}, respectively. Each panel shows the distribution of $\rho_{\rm ecl,M}$ derived from one fitted parameter.}
		\label{fig:marks_app}
	\end{figure}

	Figure~\ref{fig:marks_app} presents the outcome of this comparison. Our simulations have a physical cluster density in stars of $\rho_{\rm ecl}\approx 10^{4.5}~{\rm M_\odot\,pc^{-3}}$. For ${\cal A}^\prime_{\rm E}$, the transformed value corresponds to $\rho_{\rm ecl,M}\approx 10^{5}~{\rm M_\odot\,pc^{-3}}$, i.e.\ stronger binary disruption than expected at the actual density. This is naturally explained by the inclusion of the gas potential in our modelling, which raises the velocity dispersion and hence the encounter rate. Similarly, for ${\cal E}_{\rm cut}^\prime$, the inferred $\rho_{\rm ecl,M}$ is also higher than the physical $\rho_{\rm ecl}$, again reflecting the higher velocity dispersion. In contrast, ${\cal S}^\prime_{\rm E}$ maps to a lower $\rho_{\rm ecl,M}$, corresponding to a steeper transition; this indicates that the peak position of the operator does not shift strongly with ${\cal E}_{\rm cut}^\prime$.

	We further note that adopting the 1, 3, and 5 Myr operators from \citet{2011MNRAS.417.1684M} yields different values of $\rho_{\rm ecl,M}$. In our simulations, owing to the presence of the gas potential and the relatively short timescale ($\leq 0.6$ Myr), the clusters do not undergo significant structural evolution. In contrast, the high-density clusters modelled by \citet{2011MNRAS.417.1684M} expand substantially due to energy release from binary hardening—so-called binary burning (see their figure~4). As a result, their binary populations follow evolutionary trajectories that diverge from our result.
	
	Overall, the $\rho_{\rm ecl,M}$ mapping provides a direct, parameter-independent way to compare our results with previous models. The systematic offsets we find underscore the importance of including the gas potential in embedded-phase simulations, and highlight that the link between operator parameters and cluster density is both time-dependent and sensitive to structural evolution.
	
\end{appendix}
\end{document}